\documentclass[sigconf]{acmart}

\def\BibTeX{{\rm B\kern-.05em{\sc i\kern-.025em b}\kern-.08emT\kern-.1667em\lower.7ex\hbox{E}\kern-.125emX}}

\newcommand{\squishlist}{
	\begin{list}{$\bullet$}
		{ \setlength{\itemsep}{0pt}
			\setlength{\parsep}{3pt}
			\setlength{\topsep}{3pt}
			\setlength{\partopsep}{0pt}
			\setlength{\leftmargin}{1.5em}
			\setlength{\labelwidth}{1em}
			\setlength{\labelsep}{0.5em} } }
	
	\newcommand{\squishlisttwo}{
		\begin{list}{$\bullet$}
			{ \setlength{\itemsep}{0pt}
				\setlength{\parsep}{0pt}
				\setlength{\topsep}{0pt}
				\setlength{\partopsep}{0pt}
				\setlength{\leftmargin}{2em}
				\setlength{\labelwidth}{1.5em}
				\setlength{\labelsep}{0.5em} } }
		
		\newcommand{\squishend}{
	\end{list}  }

\usepackage{url}
\usepackage{soul}
\usepackage{url}
\usepackage[utf8]{inputenc}
\usepackage{graphicx}
\usepackage{algorithm}
\usepackage{algorithmic}
\usepackage{amsmath}
\usepackage{booktabs}
\usepackage{multirow}
\usepackage{amssymb}
\usepackage{mathrsfs}
\usepackage{amsthm}
\usepackage[normalem]{ulem}
\usepackage{soul}
\setcopyright{acmcopyright}
\begin{document}
\fancyhead{}

\copyrightyear{2019}
\acmYear{2019}
\acmConference[CIKM '19]{The 28th ACM International Conference on Information and
Knowledge Management}{November 3--7, 2019}{Beijing, China}
\acmBooktitle{The 28th ACM International Conference on Information and Knowledge
Management (CIKM '19), November 3--7, 2019, Beijing, China}
\acmPrice{15.00}
\acmDOI{10.1145/3357384.3358006}
\acmISBN{978-1-4503-6976-3/19/11}

\title{Attributed Multi-Relational Attention Network for \\ Fact-checking URL Recommendation}


\author{Di You, Nguyen Vo, Kyumin Lee}
\affiliation{
	\institution{Worcester Polytechnic Institute}
	\state{Massachusetts}
	\country{USA}
}
\email{{dyou, nkvo, kmlee}@wpi.edu}	

\author{Qiang Liu}
\affiliation{
    \institution{Alibaba Group}
    \city{Hang Zhou}
    \country{China}
}
\email{liuq0326@126.com}


\begin{abstract}
To combat fake news, researchers mostly focused on detecting fake news and journalists built and maintained fact-checking sites (e.g., Snopes.com and Politifact.com). However, fake news dissemination has been greatly promoted via social media sites, and these fact-checking sites have not been fully utilized. To overcome these problems and complement existing methods against fake news, in this paper we propose a deep-learning based fact-checking URL recommender system to mitigate impact of fake news in social media sites such as Twitter and Facebook. In particular, our proposed framework consists of a multi-relational attentive module and a heterogeneous graph attention network to learn complex/semantic relationship between user-URL pairs, user-user pairs, and URL-URL pairs. Extensive experiments on a real-world dataset show that our proposed framework outperforms eight state-of-the-art recommendation models, achieving at least 3$\sim$5.3\% improvement.
\end{abstract}

\maketitle

%    What is the problem?
%    Why is it interesting and important?
%    Why is it hard? (E.g., why do naive approaches fail?)
%    Why hasn't it been solved before? (Or, what's wrong with previous proposed solutions? How does mine differ?)
%    What are the key components of my approach and results? Also include any specific limitations.

\section{Introduction}
\begin{figure}[t]
  \centering
  \includegraphics[width=\linewidth]{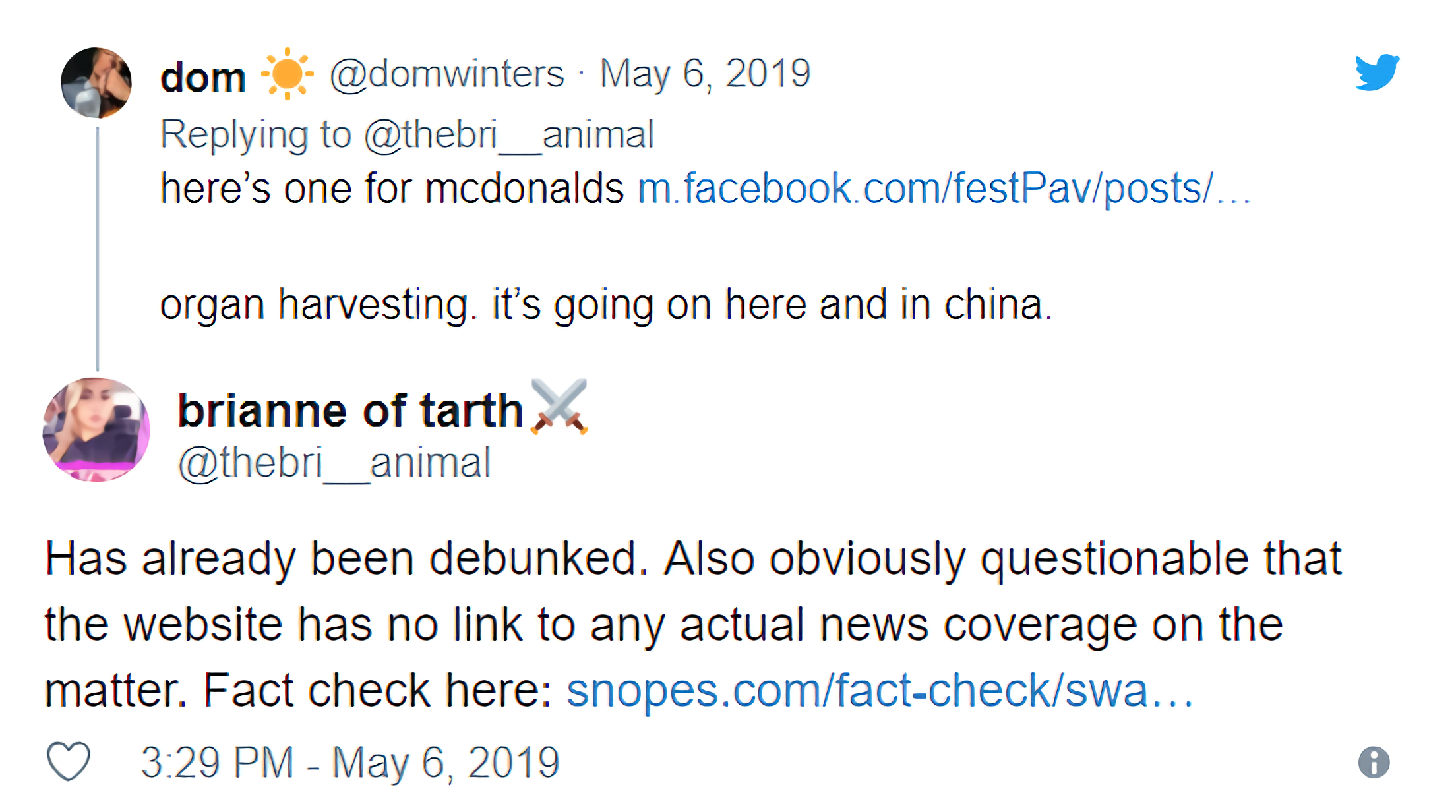}
  \caption{A real-world example of fact-checking behavior. \emph{thebri\_animal} is a fact-checker, who corrects the false claim with a fact-checking URL/article containing factual evidences.}
  \label{example}
  \vspace{-15pt}
\end{figure}

While social media sites provide users with the revolutionized communication medium by bringing the communication efficiency to a new level, they can be easily misused for widely spreading misinformation and fake news. Fake news and misinformation have been a long-standing issue for various purposes such as political propaganda \cite{allcott2017social} and financial propaganda \cite{kogan2017fake}. %They may cause more severe damage due to the nature of internet.

To fight against fake news, traditional publishers employed human editors to manually and carefully check the content of news articles to maintain their reputation. However, social media provided a new way to spread news, which lead to broader information sources and expanded audience (i.e., anyone can be a media and create news). In particular, users share news articles with their own opinion or read articles shared by their friends from whatever the source of news is with mostly blind trust \cite{twitterfake201} or with their own ideologies \cite{ecker2010explicit, nyhan2010corrections}. Although social media posts usually have a very short life cycle, the unprecedented amount of fake news may lead to a catastrophic impact on both individuals and society. Besides from misleading users with false information \cite{nyhan2010corrections}, widely propagated fake news could even cause trust crisis of entire news ecosystem \cite{Shu:2019:BeyondFN}, even further affecting both the cyberspace and physical space.

In literature, researchers focused on four topics regarding fake news: characterization (i.e., types of fake news), motivation, circulation, and countermeasures \cite{FakeNews,Zhou:2019:FNF:3289600.3291382}. A large body of work has been done on fake news identification \cite{Shu:2019:BeyondFN,Tschiatschek:2018:FND,Bastidas2018,Wang:2018:EANN} by exploiting multiple content-related and social-related components. However, we notice that the fake news still has been widely spread even after early detection \cite{figueira2017current}. Therefore, we propose to study a complementary approach to mitigate the spread and impact of fake news. Recently, community and journalists started building and maintaining fact-checking websites (e.g., Snopes.com). Social media users called \emph{fact-checkers} also started using these fact-checking pages as factual evidences to debunk fake news by replying to fake news posters. Figure \ref{example} demonstrates a real-world example of a fact-checker's fact-checking behavior on Twitter by debunking another user's false claim with a Snopes page URL as an evidence to support the factual correction.

In \cite{Vo:2018}, researchers found that these fact-checkers actively debunked fake news mostly within one day, and their replies were exposed to hundreds of millions users. To motivate these fact-checkers further quickly engage with fake news posters and intelligently consume increased volume of fact-checking articles, in this paper we propose a novel personalized fact-checking URL recommender system. According to \cite{mikolov2013distributed}, co-occurrence matrix within the given context provides information of semantic similarity between two objects. Therefore, in our proposed deep-learning based recommender system, we employ two extended matrices: user-user co-occurrence matrix, and URL-URL co-occurrence matrix to facilitate our recommendation. In addition, users tend to form relationships with like-minded people \cite{quattrociocchi2016echo}. Therefore, we incorporate each user's social context to capture the semantic relation to enhance the recommendation performance. %We also conduct extensive experiments on a real-world dataset to verify the effectiveness of our proposed frameworks.

%Based on analyzing one of the popular social media sites, Twitter, that fuels the information propagation, we propose a novel solution to harness the social media site for good toward building a personalized fact-checking article recommendation system.

%However, the analysis of fact-checking is still rudimentary. \cite{kahneman2011thinking} introduced the concept of 'What you see is all there is'(WYSIATI), indicating that people would rather rely on the information that is directly available. Accordingly, we believe that higher exposure and timely delivery of verified news article should be concerned as the first priority. Although big online social platforms begin to seek help from professional fact-checkers and develop several attempts to prevent the spread of fake news, the challenge lies that human fact-checkers cannot keep up with the volume of misinformation online which increases exponentially.

%TODO kyumin
Our main contributions are summarized as follows:
\squishlist
\item We propose a new framework for personalized fact-checking URL recommendation, which relies on multi-relational context neighbors.
\item We propose two attention mechanisms which allow for learning deep semantic representation of both a target user and a target URL at different granularity. 
\item Experimental results show that our proposed model outperforms eight state-of-the-art baselines, covering various types of recommendation approaches. Ablation study confirm the effectiveness of each component in our proposed framework.
\squishend

\section{Related Works}
In this section, we briefly review related works and position our work within the following areas: (1) fake news and misinformation; (2) advancements in recommender systems; and (3) graph convolutional networks.

\subsection{Fake News and Misinformation}
Fake news has attracted considerable attention since it is related to our daily life and has become a serious problem related to multiple areas such as politics \cite{allcott2017social} and finance \cite{kogan2017fake}. Social media sites have become one of popular mediums to propagate fake news and misinformation. The dominant line of work in this topic is fake news detection \cite{shu2017fake} which was mostly formulated as a binary classification problem. Researchers began to incorporate social context and other features for identifying fake news at an early stage and preventing it from diffusion on the social network \cite{Shu:2019:BeyondFN,Zhou:2019:FNF:3289600.3291382}. Some other researchers focus on investigating the propagation patterns of fake news in social network \cite{wu2018tracing,liu2018early}. \cite{Vo:2019:LFA} also studied fake news intervention. Unlike most previous works, we follow the direction of \cite{Vo:2018} and propose to build a personalized recommender system for promoting the fact-checking article circulation to debunk fake news.

\subsection{Advancements in Recommender System}
Traditionally, recommendation algorithms can be divided into two categories: collaborative filtering \cite{Sarwar:2001:ICF} and content-based filtering. However, in the past few years, the recommendation has become a more integrated task due to the success of the deep neural network. Neural Networks (NNs) proves to be effective to capture underlying nonlinear relations \cite{He:2017:NCF}. Another advantage is that the NNs enhanced the model's capability of extracting knowledge from multimodal data \cite{van2013deep,he2016ups,wang2017your}, which serves as auxiliary information and provide solutions to address the data sparsity problem. More recently, researchers introduced attention mechanism into recommender systems, which has achieved great success in various fields \cite{Bahdanau2015Attention,NIPS2017selfatt}. Researchers developed multiple variants of attention mechanism to improve both the recommendation precision and model interpretability \cite{wang2017dynamic, chen2017attentive, seo2017interpretable,zhu2017couplenet}.  %Attention mechanism is known for dynamically measuring confidence of each candidate’s contribution to the target object.  

In this paper, we also propose two novel designs of attention mechanism. Following \cite{Ebesu:2018:CMN,He2018NAISNA}, we further explore multi-relational context of given user-URL pair, aiming at discriminating the most important elements towards URL-dependent user preference.

%Recent advancements in neural network enhanced the prevalence of the attention mechanism.

\subsection{Graph Convolutional Networks}
With the surge of Graph-based Neural Network, GCN-based approaches have shown strong effectiveness on various tasks\cite{Kipf2016GCN,LGCL2018,NIPS2017GraphSAGE}, including recommender system. The core idea is to iteratively aggregate attributed node vectors around each node, and messages propagates by stacking multiple layers. However, the original design of GCN is not suitable for our scenario because of the following reasons: First, existing GCN works \cite{LGCL2018,NIPS2017GraphSAGE} do not distinguish different types of nodes, whereas in our case, it does not make sense to aggregate user and URL nodes together. And the aggregation function proposed in most GCN works treats all its adjacency nodes with the same importance. It is inappropriate in real-world applications and probably tends to neglect necessary information. \cite{velickovic2018gat} breaks this schema by using a multi-head attention mechanism to replace the convolution-like operator, yet it requires significant extra computation and memory. 
%Instead, we attempt to learn each user/URL node's representation as a low dimensional vector embedding in a large graph.

Compared to the previous works, in this paper, we focus on a novel application and investigate both co-occurrence context and social context related influences for fact-checking URL recommendation. We also incorporate sets of auxiliary attributes, which enable more comprehensive learning of the compatibility between given pairs of user and URL. Moreover, we take advantage of advancements in graph neural networks and attention mechanisms, and solve the aforementioned research problems.

\section{Problem Formulation} \label{definition}
We formally introduce definitions before describing our proposed framework. We define fact-checking behavior as a user (i.e., \emph{fact-checker}\footnote{We use terms user and fact-checker interchangeably in the paper.}) embeds a fact-checking URL in his reply in order to debunk fake news. We regard each fact-checking behavior as an implicit interaction between target user $i$ and target URL $j$.

\paragraph{Definition 1 (Fact-checking URL Recommendation Task)}  Let $\mathcal{U} = \{u_1,u_2,...,u_n\}$ denotes a set of fact-checkers on social media, and use $\mathcal{C} = \{c_1,c_2,...,c_m\}$ to index fact-checking URLs. We construct user-URL interaction matrix $Y = \{y_{ij} | u\in \mathcal{U}, v \in \mathcal{C} \}$ according to users' fact-checking behavior, where
\begin{equation}
    y_{ij} =
        \begin{cases}
            1, \text{if ($u_i,c_j$) interaction observed,}\\
            0, \text{otherwise.}
        \end{cases}
\end{equation}
each value of 1 for $y_{ij}$ indicates the existence of implicit interaction between target user $i$ and target URL $j$. Each user $u_i$ and each URL $c_j$ associate with a set of attributes. The goal of the recommendation task is to recommend top-N URLs from the URL set $\mathcal{C}$ to each user.

We also construct the entire dataset as a heterogeneous graph, which is a special kind of information network that consists of either multiple types of objects or different types of links, or both.

\paragraph{Definition 2 (Heterogeneous Network) \cite{sun2011pathsim}} Formally, consider a heterogeneous graph $\mathcal{G}=(\mathcal{V},\mathcal{E})$, where $\mathcal{V} (|V|= m + n)$ and $E$ denote the node set and edge set, respectively. The heterogeneity represents by the node type mapping function: $\phi: \mathcal{V} \to \mathcal{A}$ and edge type projection function: $\psi: \mathcal{E} \to \mathcal{R}$, where $\mathcal{A}$ and $\mathcal{R}$ denote the sets of predefined node types and edge types, and $|\mathcal{A}| + |\mathcal{R}| > 2$. Note that we does not consider self-loop in our graph construction.

\begin{figure}[t]
  \centering
  \includegraphics[scale=0.7]{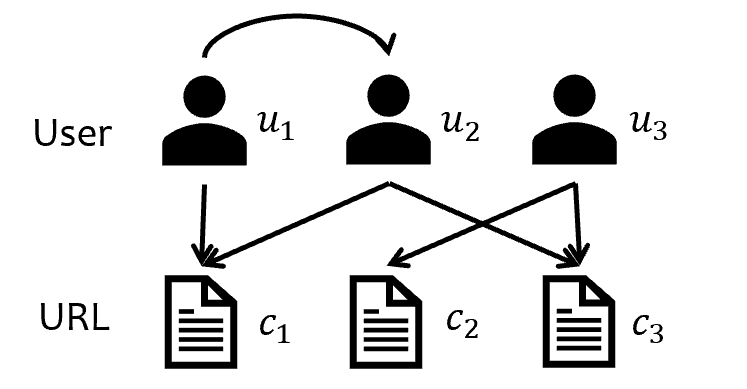}
  \caption{A toy example of multi-relational context w.r.t. given target user-URL pair.}
  \label{mr_case}
  \vspace{-16pt}
\end{figure}

\paragraph{Definition 3 (Multi-relational Context)}
Given target user $i$, we define his following fact-checkers and co-occurrenced fact-checkers as his social context user neighbors and co-occurrenced context user neighbors, respectively. Similarly, we name the other URLs posted by target user $i$ and co-occurrenced URLs of target URL $j$ as historical context URL neighbors and co-occurrenced context URL neighbors, respectively. In general, we call all the context neighbors as multi-relational context of given target user-URL pair.

\paragraph{Example} Figure \ref{mr_case} illustrates the multi-relational context. In Figure \ref{mr_case}, $c_1$, $c_2$, $c_3$ represents fact-checking URLs and $u_1$, $u_2$, $u_3$ are users who involve sharing these URLs. For example, $(u_1 \to u_2)$ indicates the social relationship between $u_1$ and $u_2$. Intuitively, we care more about the influence of $u_2$ on $u_1$. $(u_1 \to c_1 \gets u_2)$ means $u_1$ and $u_2$ are co-occurrenced user neighbors. Similarly, we name $c_1$ and $c_2$ as co-occurrenced URL neighbors of $u_3$, and $c_2$ is historical context URL neighbor given target $u_3$-$c_3$ pair.

\begin{table}[htbp]
\caption{Notations.}
\centering
\small
\begin{tabular}{@{}cl@{}}
\toprule
Notations                  & Description                                                \\ \midrule
$b_h$                      & \# of selected relation-based neighbors \\
$S$                        & Spatial weight tensor                            \\
$L$                        & Layer-wise weight tensor                         \\
$C$                        & Channel-wise wight tensor                        \\
$M$                        & Initial embedding matrix of each neighbor   \\
$N$                        & Attended embedding matrix of each neighbor  \\
$A_{ij}$                   & Weighted adjacency matrix in graph                                  \\
$W_{\phi_{i}}$             & Node type specific transformation matrix                   \\
$\mathcal{N}^{\phi_t}_i$  & Node type specific neighbor nodes                          \\
$e_{ij}^{\phi^{(l)}}$      & Importance between node pair $(i,j)$ at layer $l$                 \\
$\alpha_{ij}^{\phi^{(l)}}$ & Weights between node pair $(i,j)$ at layer $l$    \\
$p_i$                      & Neighborhood embedding of user $i$                    \\
$p_j$                      & Neighborhood embedding of URL $j$                     \\
$u^{\prime}_i$             & Wide context-based embedding of user $i$                     \\
$c^{\prime}_j$             & Wide context-based embedding of URL $j$                      \\
$h^{(l)}_i$                & Deep context-based embedding of node $i$                     \\
\bottomrule
\end{tabular}
\label{notations}
\vspace{-13pt}
\end{table}

\begin{figure*}[h]
  \centering
  \includegraphics[width=\textwidth]{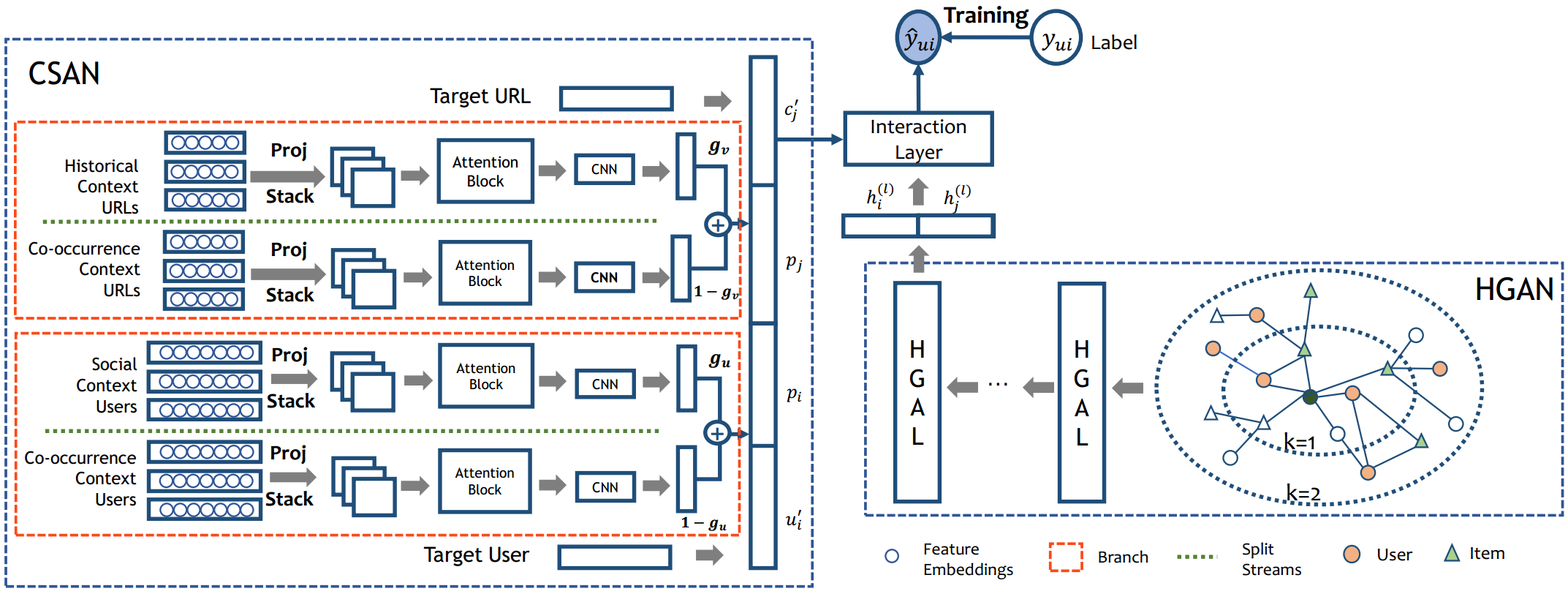}
  \caption{A schematic overview of our proposed Attributed Multi-Relational Attention Network (AMRAN), consisting of two modules: (1) a convolutional spatial attention network (CSAN); and (2) a heterogeneous graph attention network (HGAN).}
  \label{overview}
  \vspace{-15pt}
\end{figure*}

\section{Proposed Framework}
\label{sec:framework}
We propose a novel framework called Attributed Multi-Relational Attention Network (\emph{AMRAN}), to understand the influence of the multi-relational context to target user's fact-checking behavior. In this section, we elaborate our proposed AMRAN with using notations described in Table~\ref{notations}.

At the high level, \emph{AMRAN} is composed of two modules as shown in Figure \ref{overview}: (i) a convolutional spatial attention network (\emph{CSAN}) and (ii) a heterogeneous graph attention network (\emph{HGAN}). \emph{CSAN} jointly models the influence of multi-relational context on target user-URL pair (Section 4.1). It enriches the neighborhood diversity, and expands the scope of information reception. \emph{HGAN} leverages both global node connectivity and local node attributes, in order to incorporate the effect of information propagation and encode user's dynamic preference in depth (Section 4.2). At the final step, the model produces recommendations by combining wide context-aware target user embedding and URL embedding, multi-relational context user embedding and context URL embedding, and deep context-aware user embedding and URL embedding (Section 4.3).

\subsection{Convolutional Spatial Attention Network (CSAN)}
The left bounding box in Figure \ref{overview} illustrates the structure of CSAN module. To provide a broad scope of knowledge for generating wide context-aware target user embedding and URL embedding, we adopt a multi-branch setting in CSAN. The two parallel branch models multi-relational context for target user and target URL respectively. Each branch contains two identical streams. We select $b_h$ context neighbors for each stream (e.g.,  historical context URL neighbors and co-occurrenced context URL neighbors of target URL, social context user neighbors and co-occurenced user neighbors of target user). These streams are employed to learn the most discriminative features from multi-relational neighbors of target user and target URL. Then we employ a gated fusion layer to capture the optimal global level representation of target user-URL pair.

Note that we enable the embedding sharing within each branch as users/URLs share the same feature set.

\subsubsection{Raw Attribute Input}
User and URL associate with different feature sets. Therefore, CSAN starts from embedding the input attribute set of each context neighbor. We use $s$ and $t$ to denote the number of features related to user and URL, respectively. Note that the dimension of initial embedding for each attribute could be different since they may carry with different information volume. We use one-hot encoding for categorical feature inputs, and apply direct lookup on these features. However, the same solution performs poorly when it comes continuous attributes such as the post frequency of an URL. Empirically, we found that an available solution is to bucketize these features into small intervals. Specifically, we map these continuous attributes in range $[0,1), [1,2),..., [2^k, 2^{k+1})$ into $0,1,..., k$ in this work.

\subsubsection{Attribute Embedding Layer}
We then project them into the same latent space via a set of attribute-specific transformation matrices $W_1, W_2, ..., W_{s+t}$ to project all the attributes into a $w$-dimensional space. The attributes of each neighbor then are stacked as a matrix in shape of $s \times w$ for users and $t \times w$ for URLs.

However, we treat the target user-URL pair differently. After projecting attributes by the same attribute-specific transformation matrix as their relational neighbors, instead of stacking them as a matrix, we concatenate the attribute embedding vectors together and feed it through a linear projection to generate $u^{\prime}_i \in \mathbb{R}^d$ and $c^{\prime}_j \in \mathbb{R}^d$ for future reference.

\subsubsection{Spatial Attention Block}
\begin{figure}[t]
  \centering
  \includegraphics[width=\linewidth]{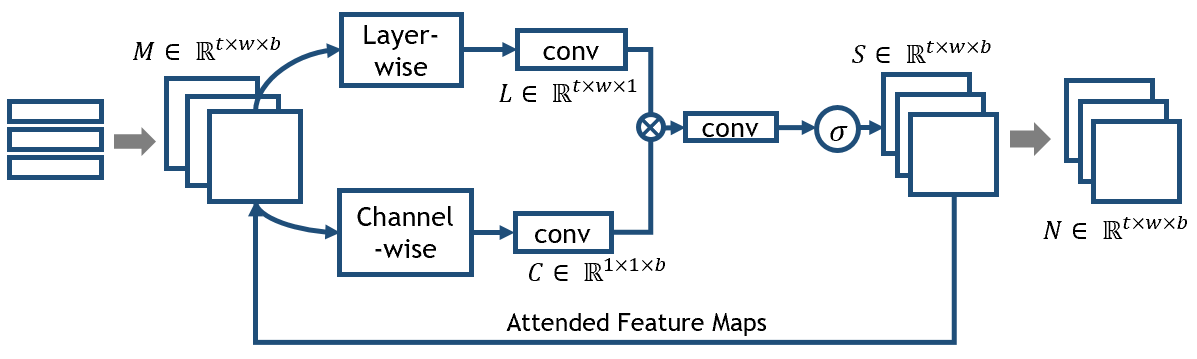}
  \caption{The illustration of Spatial Attention Mechanism (show an attention block in the historical context URL stream for illustration).}
  \label{spat_attn}
  \vspace{-15pt}
\end{figure}

To prevent some unknown misalignment and conduct better comparison among the neighborhood features, we proposed a schema for jointly learning the layer-wise and channel-wise attention. In particular, for each stream, we pile the neighbors' representation matrices together to obtain a $3$-dimensional tensor $M$. Intuitively, the design helps improve the alignment quality of neighbor's features. Then, inspired by \cite{hu2018senet,li2018harmonious}, we employ a spatial attention block in each stream for jointly learning channel-level and layer-level soft attention. See figure \ref{spat_attn} for a high-level illustration of our spatial attention block. All the streams adopt identical spatial attention blocks, and each block attends the input attribute representations independently.

In the figure, we use the historical context URL stream for illustration. The output of spatial attention block is an attention weight map $S \in \mathbb{R}^{t \times w \times b}$ which is in the same shape with the input tensor $M$. Intuitively, the layer-wise attention and channel-wise attention are dedicated to selecting the most discriminative features and the most important neighbors, respectively. Thus, they are highly complementary to each other in functionality; and we adopt a factorized manner for optimization and computational efficiency as:
\begin{equation}
    S = L \times C
\end{equation}
where $L \in \mathbb{R}^{t \times w \times 1}$ and $C \in \mathbb{R}^{1 \times 1 \times b}$ denote the layer-wise feature map and channel-wise feature map, respectively. $S$ is the result of tensor multiplication.

\paragraph{\textbf{Layer-wise Attention}}
Conceptually, the layer-wise attention learns globally important elements in the feature. We apply a cross-channel average pooling operation onto the input tensor, following by 2 convolution layers of $3 \times 3$ and $1 \times 1$ filter, respectively. Specifically, cross-channel average pooling operation is defined as:
\begin{equation}
    L = \frac{1}{b}\sum_{b^{\prime}=1}^b  M_{1:t,1:w,b^{\prime}}
\end{equation}
where $b$ is the number of selected neighbors.

\paragraph{\textbf{Channel-wise Attention}}
The design of channel-wise attention is very similar to layer-wise attention, which aims to acquire a global view of discriminative users.
Formally, the global average pooling is defined as:
\begin{equation}
    C = \frac{1}{t \times w}\sum_{w^{\prime}=1}^w \sum_{t^{\prime}=1}^t  M_{t^{\prime},w^{\prime},1:b}
\end{equation}
where $t$ and $w$ are shared height and width of all channels. Similarly, we employ two convolution layers after the pooling operation.

Note that each convolution layer was followed by batch normalization operation. Furthermore, as other work of modern CNN structure \cite{cnnrelu}, we append a ReLU activation function to assure $L>0, C>0$.

We further introduce one more convolution layer of $1 \times 1 \times b$ filter for enhancing the fusion of the layer-wise attention and channel-wise attention. The output tensor then is fed through a sigmoid function for normalization and generate the final attention weight tensor of spatial attention block. Formally, the output of the spatial attention module is the element-wise product of initial feature tensor $M$ and generated attention weights $S$:
\begin{equation}
    N = M \odot S
\end{equation}

Intuitively, the attended feature map learned fine-grained important elements via high alignment and compatible attentions.

\subsubsection{Gated Branch Fusion Layer}
We apply another CNN layer of $3 \times 3$ filter after the attended user representation of each stream for feature extraction and dimension :
\begin{equation}
    N_{op}= ReLU(W N)
\end{equation}

\begin{equation}
    p^k = MAXPOOLING(N_{op})
\end{equation}
which produces the multi-relational context representation vectors: $o_{i_h}, o_{i_c}, o_{u_f}$ and $o_{u_c}$ for each stream, respectively.

We employ a gated mechanism to assigns different weights to relation-specific neighborhood representation as:
\begin{equation}
    p_i = g_u \cdot o_{u_f} + (1 - g_u) \cdot o_{u_c}
\end{equation}

\begin{equation}
    p_j = g_v \cdot o_{i_h} + (1 - g_v) \cdot o_{i_c}
\end{equation}
where scalars $g_u$ and $g_v$ are learned automatically to control the importance of the two streams within each branch.
\begin{comment}
Furthermore, we formulate the output of this module as:
\begin{equation}
    o_m = u^{\prime}_i \oplus p_j \oplus p_i \oplus c^{\prime}_j
\end{equation}
where $\oplus$ denotes vector concatenation.
\end{comment}

\subsection{Heterogeneous Graph Attention Network (HGAN)}
Following recent success in Graph Convolutional Network (GCN) \cite{Kipf2016GCN,LGCL2018,RGCN2018,NIPS2017GraphSAGE,velickovic2018gat}. We propose a heterogeneous graph attention network (HGAN) which is tailored for recommendation task. In particular, our proposed module adopts a parallel attention structure for the user neighbor and the URL neighbor of the central node, respectively.
Considering a heterogeneous graph $\mathcal{G}=(\mathcal{V},\mathcal{E})$, the nodes represent objects in this network which can be either user or URL. The edges denote the relation between connected nodes. The node attributes pass along the edges during the propagation. We try to leverage between the local node attributes and global network structure. Our novelty lies in two aspects: (i) we differentiate the contribution of URL node and user node, respectively; and (ii) we consider both similarities of node and the influence of different relation types.

While the CSAN obtains information from multi-relational immediate neighbors, which expand the scope of knowledge for target user and target URL representations,  HGAN aims at learning deeper semantic representations of target user and target URL.

\subsubsection{Heterogeneous Graph Network}
We try to capture different semantic relation behind various types of nodes and edges. For every single layer, if the central node is user node, its neighborhood contains its co-occurrenced users and posted URLs. If the central node type is URL, its neighborhood nodes consist of users who posted it and its co-occurrenced URLs.

We adopt similar embedding approach as we did in CSAN for the initial representation of each node, but we concatenate all the features into a long vector $x_i$ for each node instead of stacking them as a matrix. Considering the different types of the node associated with the varied feature set, we use a set of node type-specific transformation matrices to project different types of node representation into the same feature space before aggregation as follows:
\begin{equation}
    h^{(0)}_i= W_{\phi_i} \cdot x_i
\end{equation}
Let $H^{(0)} \in \mathbb{R}^{(m+n) \times d}$ be the embedding matrix of all the attributed nodes, where $m+n$ is the total number of nodes and d is the dimension of latent embedding space; each row $h_i^{(0)}$ stands for the initial embedding vector of node $i$.

We define edges based on users' reference of URL (user-URL edges), user co-occurrence relation (user-user edges), and URL co-occurrence (URL-URL edges). We then introduce an adjacency matrix $A$ of $\mathcal{G}$ based on the importance of each edge. In particular, to compute the weight of user-user edges and URL-URL edges, we adopt a matrix named Shifted Positive Point-wise Mutual Information (SPPMI) \cite{NIPS2014SPPMI}, a popular measure for word associations, to utilize the co-concurrence context information. In word embedding scenario, each cell within the matrix measures the relation of corresponding word-context pair. The factorization of such matrix is proved to be equivalent to skip-gram model with negative sampling (SGNS). The Point-wise Mutual Information (PMI) between node $i$ and node $j$ is computed as $PMI(i,j) = log \frac{P(i,j)}{P(i)P(j)}$ where $P(i,j) = \frac{\# (i,j)}{|D|}$ and $P(i) = \frac{\# (i)}{|D|}$. $|D|$ denotes the total number of observed word-context pairs within a predefined sliding window. $P(i,j)$ is the joint probability that word $i$ and word $j$ appear together within the window size. Furthermore, we introduce the SPPMI matrix as an extension based on PMI value:
\begin{equation}
    SPPMI(i,j)=max\{PMI(i,j)-log(k),0\}
\end{equation}
where $k$ is a hyperparameter, which represents the number of negative samples. Conceptually, a positive PMI value implies a semantically correlated word-context pair, Therefore, SPPMI, which only takes the positive value of PMI shifted by a global constant, reflects a closer semantic relation between word-context pairs. Inspired by this concept/idea, we use $|D|$ to denote the number of times of user (URL) co-occurrence and generate the user co-occurrence matrix in shape of $n \times n$ and URL co-occurrence matrix of $m \times m$. Note that we do not discriminate between the target node and context node.

Similarly, we learn from the TF-IDF concept and redefine it on recommendation task with implicit feedback \cite{fayyadadvances} as:
\begin{equation}
    TF-IDF_{ij} = TF_{ij} \times IDF_i = \frac{\# (i,j)}{\max_k \# (i,k)} log \frac{m}{m_i}
\end{equation}
where $\# (i,j)$ represents the number of times URL $j$ be posted by user $i$. $TF_{ij}$ further normalizes it by the maximum number of post times of any URL by user $i$. The $IDF_i$ is associated with the user's previous behavior as $m$ denotes the total number of URLs and $m_i$ is the number of URLs posted by user $i$.

Formally, the weight of the edge between node $i$ and node $j$ is defined as:
\begin{equation}
    A_{ij} =
        \begin{cases}
            SPPMI(i,j) & \text{$i,j$ are user (URL)}\\
            TF-IDF_{ij} & \text{$i$ is user, $j$ is URL}\\
            1 & \text{i=j,}\\
            0 & \text{otherwise}
        \end{cases}
\end{equation}

\subsubsection{Heterogeneous Attention Layer (HGAL)}
Given the node's initial representation defined as above, we then pass messages to aggregate the neighborhood nodes' information and combine it with the target user's interests. A popular propagation strategy in existing GCN works is the normalized Laplacian matrix \cite{Kipf2016GCN}. Even though it proves to be effective, it is not trainable and it assigns every adjacent node with the same weight. Following previous work \cite{velickovic2018gat}, we propose to incorporate a hierarchical attention mechanism to learn the weight of each adjacent node adaptively.

\begin{figure}[t]
  \centering
  \includegraphics[width=\linewidth]{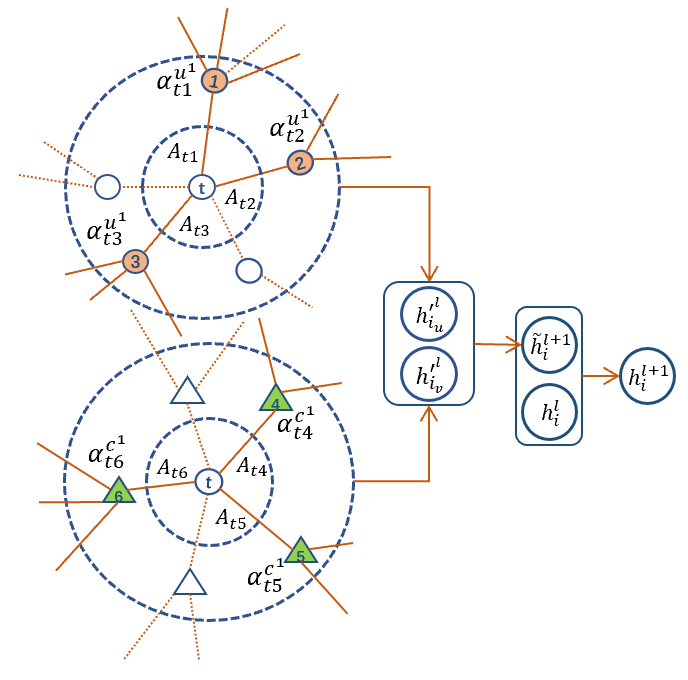}
  \caption{Graphical illustration of a single heterogeneous graph attention layer. In this example, we assume the central node as a user node. Circles denote users, and triangles denote URLs. Colored objects with a solid line are selected neighbors at each layer, and the nodes with a dotted line are randomly dropped. (Best viewed in color).}
  \label{hgat}
  \vspace{-15pt}
\end{figure}

Since the distribution of the number of neighbors of each node disperses greatly, sub-sampling becomes an essential procedure in our task to avoid an explosion of computation cost after multiple hops stacked. We adopt Weighted Random Selection (WRS) \cite{efraimidis2006WRS} to select a fixed number of nodes for both node types in each graph attention layer. Figure \ref{hgat} shows a graphical illustration of one HGAL.

Assume that the central node is a user node. We separately calculate the attention weights between the user node and its user node neighbors, or between the user node and its URL node neighbors. The similarity between the target user's node representation $h^{(l)}_u$ and all of its selected neighbors are defined as:
\begin{equation}
    \alpha_{ij}^{\phi^{(l)}} = softmax(e_{ij}^{\phi^{(l)}}) = \frac{exp(f(h^{(l)}_i,h^{(l)}_j))}{\sum_{k \in \mathcal{N}^{\phi_t}_i} exp(f(h^{(l)}_i,h^{(l)}_k))}
\end{equation}
where $h^{(l)}_i$ is the representation of user $i$ at layer $l$, and $\mathcal{N}^{\phi_t}_i$ denotes the node type-based neighbor. We adopt $f(h^{(l)}_i,h^{(l)}_j)=cosine(h^{(l)}_i,h^{(l)}_j)$ as similarity function. Intuitively, $\alpha^{\phi}_{ij}$ measures the importance of neighbor $j$ towards central node $i$. Meanwhile, we obtain the edge weight $A_{ij}$ as well.

After this, we aggregate the type-based neighborhood node representation and generate the embedding of neighborhood as the average of different types of nodes:
\begin{equation}
    z_{ij} = ReLU(A_{ij} h^{(l)}_i)
\end{equation}

\begin{equation}
    \tilde{h}^{(l+1)}_i = \frac{1}{|\mathcal{A}|}(\sum_{j \in \phi_{\mathcal{U}}} \alpha_{ij}^{\phi^{(l)}} z_{ij} + \sum_{j \in \phi_{\mathcal{C}}} \alpha_{ij}^{\phi^{(l)}} z_{ij})
\end{equation}

To model the information propagation and capture higher-order relations, we stack the HGAL multiple times. In addition, we introduce the residual connection \cite{He2016skipconnection} to help train a HGAN with many layers.
\begin{equation}
 g^{(l+1)} = \sigma(W_g^{(l)} h^{(l)} + b_g^{(l-1)})
\end{equation}

\begin{equation}
  h^{(l+1)}= (1 - g^{(l+1)}) \odot \tilde{h}^{(l+1)}_i + g^{(l+1)} \odot h^{(l)}
 \end{equation}
where $\sigma$ denotes the sigmoid function. $W_g^{(l)}$ and $b_g^{(l-1)}$ are the shared weight matrix and bias term at layer $l$, respectively. The node representation at $l$-th layer provides knowledge of $l$ degrees away.

\subsection{Interaction Layer}
The interaction layer is tailored for recommendation tasks. Recall that we obtained wide context-based user embedding $u^{\prime}_i$ and URL embedding $c^{\prime}_j$, context representations $p_i$, $p_j$ and deep context-based user embedding $h^{(l)}_i$ and URL embedding $h^{(l)}_j$ in the previous sections. Then we formulate the final URL-dependent user representation by using a fully connected layer as:
\begin{equation}
    o_i = W_o[u^{\prime}_i \oplus c^{\prime}_j \oplus p_i \oplus p_j \oplus h^{(l)}_i \oplus h^{(l)}_j] + b_o
\end{equation}
where $W_o$ and $b_o$ are a linear transformation weight matrix and bias term, respectively. $\oplus$ denotes vector concatenation. Note that the fully-connected layer can be replaced by other techniques (e.g. CNN). Finally, we feed it through a softmax function to calculate the probability that user interested in the given URL.

\subsection{Training}
We adopt the cross-entropy loss function during the training process.
\begin{equation}
\mathcal{L} = - \sum_{(i,j) \in Y^{+} \bigcup Y^{-}} y_{ij}log (\hat{y}_{ij}) + (1 - y_{ij}) log (1 - \hat{y}_{ij})
\end{equation}
We follow a uniform sampling strategy to obtain negative samples $(i,j) \in Y^{-}$ from unobserved interactions.
Since the entire architecture is differentiable, we use back propagation to achieve end-to-end training.

\section{Evaluation}
In this section, we describe a dataset, baselines, experimental setting, and experimental results. In the experiments, we seek to answer the following research questions:
\squishlist
\item \textbf{RQ1:} What is the performance of our model and baselines?
\item \textbf{RQ2:} How beneficial is each submodule of our model?
\item \textbf{RQ3:} How effective is our attention mechanisms?
\item \textbf{RQ4:} What is sensitivity of our model with regard to hyperparameters?
\squishend

%FCRG-DT is based on the dot attention whereas FCRG-BL is based on the bilinear attention described in Eq. \ref{eq:attention_mechanism}.

%We highlight the following results which demonstrate the effectiveness of our proposed model.
%\begin{itemize}
%    \item[-] Our proposed AMRAN model outperforms 7 state-of-art baselines which cover a different type of recommendation methods.
%    \item[-] We conduct ablation tests to confirm the usefulness of the different modules in our architecture.
%\end{itemize}

\subsection{Dataset}
We evaluate our proposed model on a Twitter dataset obtained from the authors of \cite{Vo:2018}\footnote{https://github.com/nguyenvo09/CombatingFakeNews}. The interaction behavior collected in the dataset is consistent with our definition in \ref{definition}. As they did for their study, we only kept users who have at least three interactions (i.e., posting at least three fact-checking messages containing fact-checking URLs). We conducted additional preprocessing step by removing users whose posts are non-English, or their tweets were inaccessible, because some of our baselines require a fact-checker's tweets. Our final dataset consists of 11,576 users (i.e, fact-checkers), 4,732 fact-checking URLs and 63,429 interactions.
The dataset also contains each user's social network information. Note that each user's social relationship is restricted within available users in the dataset. And we further take available feature values of both user and URL into consideration. For instance, a category of referred fact-checking article and the name of corresponding fact-checking website reveals linguistic characteristics such as writing style and topical interest of each URL; while the number of followers and number of followees of each user indicates the credibility and influence of the fact-checker. Statistics of the final dataset is presented in Table \ref{data_stat}.

\subsection{Baselines}
To measure relative effectiveness of our model, we compare our model against eight state-of-the-art baselines including the traditional collaborative filtering method, neural network-based models, and context-aware approaches.

\squishlist
\item \textbf{MF} \cite{Koren:2009:MFT} is a standard collaborative filtering technique. It factorizes an interaction matrix $X \in \mathbb{R}^{M \times N}$ into two matrices $U \in \mathbb{R}^{M \times d}$ and $X \in \mathbb{R}^{d \times N}$. $U$ contains each user's latent representation, and $X$ contains each URL's latent representation.
\item \textbf{GAU} \cite{Vo:2018} is a framework specifically designed for fact-checking URL recommendation utilizing rich side information such as a user' social network, tweets, and referred fact-checking pages. It is the most relevant and domain-specific baseline.
\item \textbf{NeuMF} \cite{He:2017:NCF} is a neural network based item recommendation algorithm. We adopted a composite version of MF jointly coupled with a MLP.
\item \textbf{CMN} \cite{Ebesu:2018:CMN} combines a global latent factor model with an augmented memory network to capture personalized neighbor-based structure in a non-linear fashion.
\item \textbf{NAIS} \cite{He2018NAISNA} is an item-based collaborative filtering architecture that integrates attention mechanism to distinguish the contribution of previously consumed items. The authors proposed two versions of NAIS: (1) $NAIS_{concat}$ which concatenates two vectors to learn the
attention weight; and (2) $NAIS_{prod}$ which feeds the element-wise product of the two vectors to the attention network. Therefore, we also build two versions of NAIS, and compare them with our model.
\item \textbf{DeepCoNN} \cite{zheng2017joint} was originally proposed for an item rating prediction task which jointly model user and item based on their textual reviews. The prior work shows that it significantly outperforms other topic modeling based methods.We re-implemented the baseline and adapted it for our recommendation task with implicit feedback.
\item \textbf{NARRE} \cite{chen2018neural} is a deep neural network based framework for a item rating prediction task. It employs the attention mechanism to distinguish the importance of each review. We re-implemented the framework for our implicit feedback situation.
\item \textbf{NGCF} \cite{NGCF19}  is a new recommendation framework based on graph neural network, explicitly encoding the collaborative signal in the form of high-order connectivity in user-item bipartite graph by performing embedding propagation.
\squishend

Table~\ref{approaches} presents characteristics of baselines and our model, showing what information each model utilizes. Note that even though CMN and NAIS both utilize co-occurrence context, CMN only utilizes user co-occurrence context whereas NAIS looks into URL co-occurrence context.

\begin{table}[t]
\caption{Statistics of our evaluation dataset.}
\centering
\small
\begin{tabular}{@{}llll@{}}
\toprule
Interaction \# & User \# & URLs \# & Sparsity \\ \midrule
63429         & 11576  & 4732   & 99.884\%  \\ \bottomrule
\end{tabular}
\label{data_stat}
\vspace{-13pt}
\end{table}

\begin{table*}[th]
\caption{Characteristics of baselines and our model.}
\centering
\small
\begin{tabular}{@{}lccccccccc@{}}
\toprule
                      & MF          & GAU         & NeuMF       & CMN         & NAIS        & DeepCoNN    & NARRE       & NGCF       & AMRAN       \\ \midrule
Implicit Feedback     & $\surd$     & $\surd$     & $\surd$     & $\surd$     & $\surd$     & $\surd$     & $\surd$     & $\surd$     & $\surd$ \\
Textual Content       & $\setminus$ & $\surd$     & $\setminus$ & $\setminus$ & $\setminus$ & $\surd$     & $\surd$     & $\setminus$      & $\setminus$ \\
Co-occurrence Context & $\setminus$ & $\surd$     & $\setminus$ & $\surd$     & $\surd$     & $\setminus$ & $\setminus$ & $\setminus$ & $\surd$     \\
Social Context        & $\setminus$ & $\surd$     & $\setminus$ & $\setminus$ & $\setminus$ & $\setminus$ & $\setminus$  & $\setminus$ & $\surd$     \\
Higher-order Information        & $\setminus$ & $\setminus$     & $\setminus$ & $\setminus$ & $\setminus$ & $\setminus$ & $\setminus$  & $\surd$ & $\surd$     \\
Deep Learning         & $\setminus$ & $\setminus$ & $\surd$     & $\surd$     & $\surd$     & $\surd$     & $\surd$     & $\surd$       & $\surd$     \\ \bottomrule
\end{tabular}
\label{approaches}
\vspace{-13pt}
\end{table*}

%\subsection{Experiment details}
\subsection{Evaluation Protocol}
We adopt the leave-one-out evaluation protocol to evaluate the performance of our model and baselines. The leave-one-out evaluation protocol has been widely used in top-K recommendation tasks. In particular, we held the latest interaction of each user as the test set and used the remaining interactions for training. Each testing instance was paired with 99 randomly sampled negative instances. Each recommendation model ranks the 100 instances according to its predicted results. The ranked list is judged by Hit Ratio (HR) \cite{Deshpande:2004:HR} and Normalized Discount Cumulative Gain (NDCG) \cite{He:2015:NDCG} at the position 10. HR@10 is a recall-based metric, measuring the percentage of the testing item being correctly recommended in the top-10 position. NDCG@10 is a ranked evaluation metric which considers the position of the correct hit in the ranked result. Since both modules in our framework introduce randomness, we repeat each experiment 5 times with different weight initialization and randomly selecting neighbors. We report the average score of the best performance in each training process for both metrics to ensure the robustness of our framework.

\subsection{Hyper-parameter Settings}
We implement our framework by using Pytorch framework, initialize weight parameters by Xavier initialization \cite{Goodfellow-et-al-2016}, and optimize the model with Adam optimizer \cite{kingma2014adam}. The mini-batch size is set to 128. Empirically, in CSAN, we select 10 neighbors for each stream. In HGAN, we choose 8 user neighbors and 8 URL neighbors for each central node at a single layer, and the default number of graph attention layers is set to 2. If the object (i.e.g, user neighbor or URL neighbor) is not sufficient enough, we pad the sequence with zeros vectors.

In the proposed AMRAN model, all hyperparameters are tuned by using the grid-search on the validation set, which is formed by holding out one interaction of each user from the training data like the prior work \cite{He:2017:NCF}. We conduct the grid search over a latent dimension size from \{8,16,32,64\}, a regularization term from \{0.1, 0.01, 0.001, 0.0001, 0.00001\}, a learning rate from \{0.0001, 0.0003, 0.001, 0.01, 0.05, 0.1\}, and SPPMI shifted constant value $s$ from \{1, 2, 5, 10\}. The number of negative samples w.r.t each positive interaction is set to 4. We adopt the same latent dimension size for all sub-modules. For a fair comparison, we also thoroughly optimize the baselines' hyperparameters by using the validation set.
%kyumin plateaued??

\subsection{RQ1: Performance of Our Model and Baselines}

\begin{table}[htbp]
\caption{Performance of our AMRAN and baseline models. AMRAN outperforms all baselines in both evaluation metrics.}
\centering
\small
\begin{tabular}{@{}lll@{}}
\toprule
Model   & HR@10  & NDCG@10 \\ \midrule
MF & 0.537 & 0.364\\
GAU       & 0.589 & 0.372  \\
NeuMF       & 0.621 & 0.389  \\
CMN         & 0.589 & 0.382  \\
NAIS\_prod & 0.617 & 0.392  \\
NAIS\_concat & 0.624 & 0.398  \\
DeepCoNN         & 0.609 & 0.377  \\
NARRE & 0.615 & 0.382 \\
NGCF         & 0.600 & 0.373  \\ \midrule
%\emph{our} CSAN  & 0.642          & 0.387          \\
%\emph{our} HGAN  & 0.653          & 0.403          \\
\emph{our} AMRAN & \textbf{0.657} & \textbf{0.410} \\
\bottomrule
\end{tabular}
\label{results}
\vspace{-13pt}
\end{table}

Table~\ref{results} presents performance of our model and baselines. According to the results and information described in Table \ref{approaches}, we had the following observations. First, deep learning-based approaches usually obtained better performance than traditional models (e.g., MF and GAU). This observation makes sense because (1) traditional models failed to capture the important non-linear relationship between users and fact-checking URLs; (2) Most deep-learning based baseline models employ attention mechanism which helps better understand the semantic relation between user and URL; and (3) training tricks such as drop out and batch normalization also contribute to a better quality of training. In particular, $NAIS_{concat}$ achieves better performance than $NAIS_{prod}$ which supports the reason (1).

%our dataset is different from general recommendation task (i.e. basket recommendation) that each user only refer to an URL once, thus degrading the performance of models which prefers to recommend previous posted URL. And

The second observation is that models with text review achieve better results compared with collaborative filtering-based methods. It is not surprising since that textual content contains rich information which could be auxiliary information to implicit feedback data and thus improve the recommendation accuracy. However, we observed that text-based recommendation approaches usually have a high complexity. Third, social context and co-occurrence context play important roles in improving recommendation results. NAIS significantly outperforms CMN and becomes the strongest baseline model. It indicates that URL-URL co-occurrence relationship is more important than user-user co-occurrence relationship since semantic representation of each user is much complex than semantic representation of a fact-checking URL.

Overall, our AMRAN outperforms all baselines, achieving 0.657 HR@10 and 0.410 NDCG@10. It improves HR@10 by 5.3\% and NDCG@10 by 3\% over the best baseline (i.e., $NAIS_{concat}$).

% In our model, we proposed two attention-based mechanisms which allow to model both user and URL semantic at the different granularity and potentially increase the performance.

\begin{table}[htbp]
\caption{Performance of two submodules (CSAN and HGAN), and AMRAN.}
\centering
\small
\begin{tabular}{@{}lll@{}}
\toprule
Model   & HR@10  & NDCG@10 \\
\midrule
\emph{our} CSAN  & 0.642          & 0.387          \\
\emph{our} HGAN  & 0.653          & 0.403          \\ \midrule
\emph{our} AMRAN & \textbf{0.657} & \textbf{0.410} \\
\bottomrule
\end{tabular}
\label{results2}
\vspace{-13pt}
\end{table}

%\subsection{Quantitative Analysis}
\subsection{RQ2: Effectiveness of our submodules}
In this experiment, we are interested in measuring effectiveness of our submodules of AMRAN: CSAN and HGAN. Table~\ref{results2} the experimental result. CSAN achieves 0.642 HR@10 and 0.387 HR@10, whereas HGAN achieves 0.653 HR@10 and 0.403 NDCG@10. Both of the submodules outperform all the baselines in HR@10. HGAN outperforms all the baselines, and CSAN is competitive over the baselines. This experimental result confirms that both CSAN and HGAN positively contributed to the performance of our AMRAN.

%We compare the entire AMRAN against its two submodules to prove the effectiveness of both components. CSAN learns from immediate multi-relational neighbors while HGAN modeling from the perspective of higher-order message dissemitation. The entire architecture make recommendation based on combining the deep and wide context-based representation together. Note that when setting the number of HGAN layer to 1, it acquires same knowledge as CSAN component excpet for users' social network. See the last three lines in Table \ref{results} demonstrate results of ablation study. Performance of overall architecture is better than two submodules, and HGAN outperforms the CSAN implies the higher-order neighbors can also have impact on user (URL) representations.

\subsection{RQ3: Effectiveness of our Attention Mechanisms}
We proposed two attention mechanisms: (1) spatial attention block in CSAN; and (2) graph attention mechanism in HGAN described in Section~\ref{sec:framework}. In this experiment, we are interested in studying the impact of the attention mechanisms. In particular, we run each submodule of AMRAN (i.e., CSAN or HGAN) with/without a corresponding attention mechanism. Table \ref{csa} shows performance of these models. In both submodules, our proposed attention mechanisms positively improved the performance of these submodules, confirming the positive impact toward correctly recommending fact-checking URLs.

\begin{table}[t]
\caption{Performance of submodules with/without our proposed attention mechanisms.}
\centering
\small
\begin{tabular}{@{}lll@{}}
\toprule
                                & HR@10 & NDCG@10 \\ \midrule
Without Spatial Attention Block &  0.614     &  0.368     \\
CSAN                            &   0.642   &  0.387      \\ \midrule
Without Graph Attention Mechanism &  0.638     &  0.389       \\
HGAN                            &    0.653   &  0.403   \\ \bottomrule
\end{tabular}
\label{csa}
\vspace{-13pt}
\end{table}

\subsection{RQ4: Hyperparameter Sensitivity}
\begin{figure}[t]
  \centering
  \includegraphics[width=\linewidth]{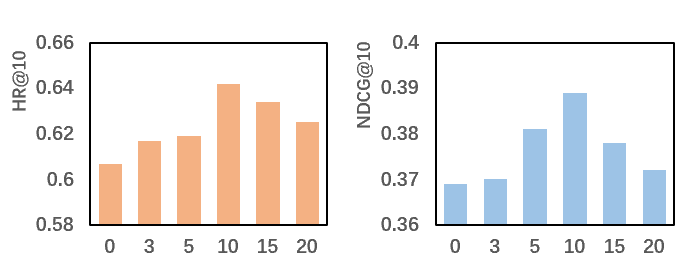}
  \caption{Performance of CSAN when varying the number of neighbors in each stream.}
  \label{csan_exp}
  \vspace{-15pt}
\end{figure}

\begin{figure}[t]
  \centering
  \includegraphics[width=\linewidth]{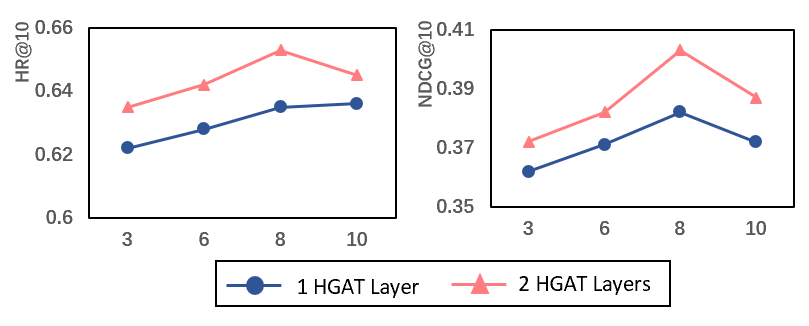}
  \caption{Performance of HGAN when varying a size of neighbor nodes at each layer (HGAL).}
  \label{hgat_exp}
  \vspace{-15pt}
\end{figure}

\begin{figure}[t]
  \centering
  \includegraphics[width=\linewidth]{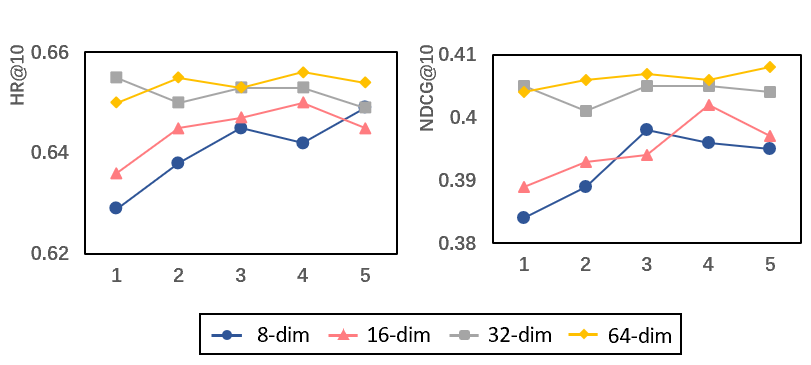}
  \caption{Performance of AMRAN when varying the number of negative samples and the size of latent semantic space (i.e., embedding size).}
  \label{hyper_exp}
  \vspace{-15pt}
\end{figure}
Now, we turn to analyze how our model is sensitive to hyperparameter values, and which hyperparameter value produces the best recommendation result. Recall that we utilize the context information to generate comprehensive embedding of given user and URL. In CSAN, we employ four streams to capture fine-grained context characteristics and share the embedding weight matrix with the target user and target URL representations. In the first experiment, we vary the number of neighbors associated with each steam in CSAN to show how CSAN's performance is changed. Figure \ref{csan_exp} shows that both $HR@10$ and $NDCG@10$ have similar trends, and selecting 10 neighbors at each stream produced the best result.

Next, we measure how performance of HGAN is changed when varying the number of HGALs and a size of selected neighbor nodes at each layer. Figure \ref{hgat_exp} demonstrates the necessity of employing 2 HGALs, which consistently outperforms the one HGAL. The best performance was achieved when a size of selected neighbor nodes was set to 8. In addition, we vary the number of negative samples, and a size of latent semantic space for the target user and target URL (i.e., an embedding vector size of the target user and target URL). Figure \ref{hyper_exp} shows high dimensional latent semantic space produces high performance of AMRAN. 64 dimensional embeddings produced the best results. We also observe that one negative sample would not be enough to produce good results in especially when an embedding vector size is small. The top performance is achieved when one positive instance paired with 3 or 4 negative instances.

%how embedding size affects the performance of ourthe impact of more general parameters. In particular, we experiment with
%. Figure \ref{hyper_exp} suggests that lower dimension of latent space fails to convey the complete semantic. However, the improvement is marginal from 32-dimensional to 64-dimensional latent space. We also observe that 1 negative sample would introduce much too randomness into the training process, and embedding vectors with high dimension are less sensitive to the number of negative samples. The top performance is when one positive instance paired with 3~4 negative ones.

\subsection{Case Study: Visualization of Relevance Propagation}

\begin{figure}[t]
  \centering
  \includegraphics[width=\linewidth]{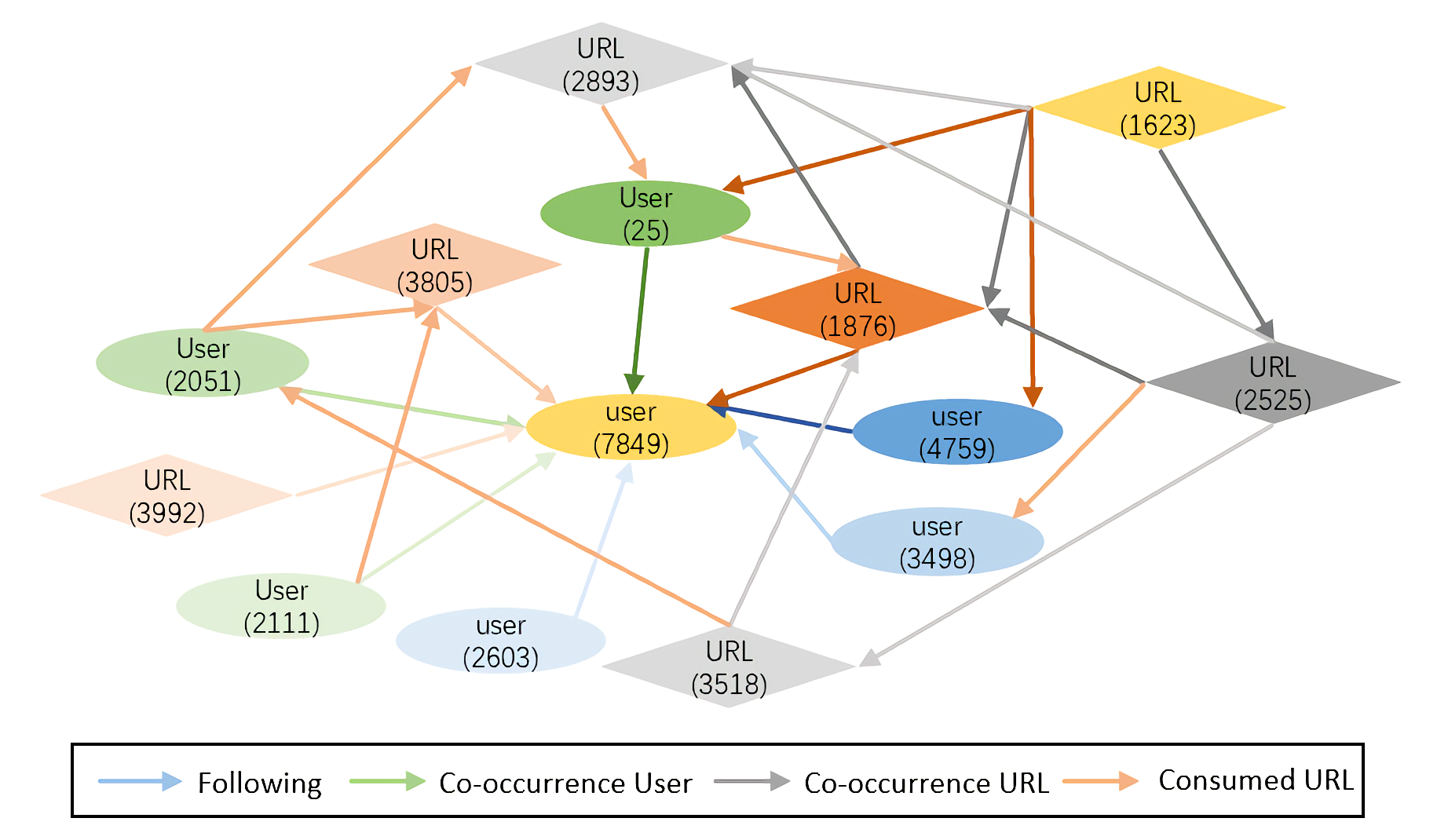}
  \caption{Visualization of relevance propagation of a user 7849. Objects in yellow denote target user and target URL. (Best viewed in color).}
  \label{case}
  \vspace{-15pt}
\end{figure}

Attention mechanism not only improve recommendation performance of our model, but also provide explainability of our model. As a case study, we specifically chose an example to demonstrate relevance propagation. In particular, we randomly sampled a user 7849 as the example as shown in Figure \ref{case}. The user 7849 has 3 co-occurrenced users, 3 following users, and posted 4 URLs. Note that we omit less important 2nd-degree neighbors for simplicity. The most relevant neighbors and the propagation paths are highlighted automatically via the attention mechanism. In general, based on the user's historical context URLs, we observe that the topic that user 7849 would like to participate in debunking is fauxtography. However, in this very particular case, the most influential context neighbors of the user are user 25 (co-occurrence user) and user 4759 (social context) given URL 1623. Both of the context neighbors share the similar taste with user 7849 on the favorite website (Politifact.com). Moreover, we found that URL 2525 appeared in 2nd-degree neighborhood of the user 7849, and was originated from the same website (Snopes.com) with URL 1623.

\section{Conclusion}
In this paper, we proposed a novel framework, which effectively recommends relevant fact-checking URLs to \emph{fact-checkers}. The proposed framework inspired by recent advancements in graph neural network and attention mechanism leveraged user-URL specific context information to capture deep semantic and complex structure between target user and target URL. We compared the performance of our model, AMRAN, with eight state-of-the-art baselines. Experimental results showed that our model achieved up to 5.3\% improvement against the best baseline. Both submodules of AMRAN positively contributed to the recommendation results. %In the future, we are interested in generalizing our proposed method for general recommendation tasks, and incorporate other auxiliary information such as immense text information and temporal signals to further improve performance of our model.

\begin{acks}
This work was supported in part by NSF grant CNS-1755536, AWS Cloud Credits for Research, and Google Cloud. Any opinions, findings and conclusions or recommendations expressed in this material are the author(s) and do not necessarily reflect those of the sponsors.
\end{acks}

\bibliographystyle{ACM-Reference-Format}
\bibliography{ref}

\appendix

\end{document}